# Multi-GPU implementation of a VMAT treatment plan optimization algorithm

Zhen Tian[1], Fei Peng[2], Michael Folkerts[1], Jun Tan[1], Xun Jia[1] and Steve B. Jiang[1]

[1]Department of Radiation Oncology, University of Texas, Southwestern Medical Center, Dallas, TX 75390

[2]Computer Science Department, Carnegie Mellon University, Pittsburgh, PA 15213

Emails: Zhen.Tian@UTSouthwestern.edu, Xun.Jia@UTSouthwestern.edu, Steve.Jiang@UTSouthwestern.edu

**Purpose**: VMAT optimization is a computationally challenging problem due to its large data size, high degrees of freedom, and many hardware constraints. High-performance graphics processing units (GPUs) have been used to speed up the computations. However, GPU's relatively small memory size cannot handle cases with a large dose-deposition coefficient (DDC) matrix in cases of, e.g., those with a large target size, multiple targets, multiple arcs and/or small beamlet size. The main purpose of this paper is to report an implementation of a column-generation based VMAT algorithm, previously developed in our group, on a multi-GPU platform to solve the memory limitation problem. While the column-generation based VMAT algorithm has been previously developed, the GPU implementation details have not been reported. Hence, another purpose is to present detailed techniques employed for GPU implementation. We also would like to utilize this particular problem as an example problem to study the feasibility of using a multi-GPU platform to solve large-scale problems in medical physics.
**Methods:** The column-generation approach generates VMAT apertures sequentially by solving a pricing problem (PP) and a master problem (MP) iteratively. In our method, the sparse DDC matrix is first stored on CPU in coordinate list format (COO). On the GPU side, this matrix is split into four sub-matrices according to beam angles, which are stored on four GPUs in compressed sparse row (CSR) format. Computation of beamlet price, the first step in PP, is accomplished using multi-GPUs. A fast inter-GPU data transfer scheme is designed using peer-to-peer (P2P) access. The remaining steps of PP and MP problems are implemented on CPU or a single GPU due to their modest problem scale and computational loads. Barzilai and Borwein (BB) algorithm with subspace step scheme is adopted here to solve the MP problem. A head and neck (H&N) cancer case was used to validate our method. We also compare our multi-GPU implementation with three different single GPU implementation strategies: truncating DDC matrix (S1), repeatedly transferring DDC matrix between CPU and GPU (S2), and porting computations involving DDC matrix to CPU (S3), in terms of both plan quality and computational efficiency. Two more H&N patient cases and three prostate cases were also used to demonstrate the advantages of our method.

**Results:** Our multi-GPU implementation can finish the optimization process within ~1 minute for the H&N patient case. S1 leads to an inferior plan quality although its total time was 10 seconds shorter than the multi-GPU implementation due to the reduced matrix size. S2 and S3 yield the same plan quality as the multi-GPU implementation but take ~4 minutes and ~6 minutes, respectively. High computational efficiency was consistently achieved for the other 5 patient cases tested, with VMAT plans of clinically acceptable quality obtained within 23~46 seconds. Conversely, to obtain clinically comparable or acceptable plans for all these 6 VMAT cases that we have tested in this paper, the optimization time needed in a commercial TPS system on CPU was found to be in an order of several minutes.

**Conclusions:** The results demonstrate that the multi-GPU implementation of our column-generation based VMAT optimization can handle the large-scale VMAT optimization problem efficiently without sacrificing plan quality. Our study may serve as an example to shed some light on other large-scale medical physics problems that require multi-GPU techniques.

**Key words:** multi-GPU, VMAT optimization, column-generation approach

## 1. Introduction

Volumetric modulated arc therapy (VMAT) delivers a radiation treatment during continuous gantry rotations around a patient. At the same time, beam fluence map is modulated by a multi-leaf collimator (MLC)[1, 2] to yield a carefully sculpted 3D dose distribution conformal to the cancer target[3-6]. Compared with conventional Intensity Modulated Radiation Therapy (IMRT), the treatment plan optimization problem for VMAT is much more complicated. On the algorithm side, the increased degrees of freedom substantially enlarge the problem complexity. In addition to optimizing with respect to fluence map or MLC leaf positions in a typical IMRT problem[7], other linac parameters such as dose rate and gantry rotation speed may also be optimized in principle to increase the freedom to achieve an optimal solution. Moreover, these parameters are subject to strong hardware constraints, which have to be satisfied by a treatment plan to ensure its deliverability. Over the years, a number of research efforts have been devoted to developing novel algorithms to solve the VMAT optimization problem[8-24]. In particular, we have previously developed a column-generation approach[16, 18], and our test results on real patient cases have demonstrated the potential of this method in terms of generating a plan with all the degrees of freedom considered.

One challenge for VMAT optimization problems comes from the large data size. Instead of delivering the radiation beam at only a few beam angles in IMRT, VMAT beam comes along all directions in the arcs. As a consequence, a dose-deposition coefficient (DDC) matrix corresponding to all of these beam angles is needed for the optimization. As repeated forward dose calculation operations involving this huge DDC matrix are needed during an optimization process, the overall computation time of VMAT optimization is prolonged. At present, with the utilization of a multi-core computer or a computer cluster, it is possible to achieve a computation time of several minutes to tens of minutes[5, 6]. However, the overall planning efficiency is still low in clinical practice, as a trial-and-error process is usually needed to manual tune some parameters repeatedly for the optimization until achieving a good plan. Besides, sometimes several iterations between a planner and a physician are also needed, through which the planner implements the physician's intent and gradually improves the plan quality.

Recently, it has been reported that high-performance graphics processing units (GPUs) are able to significantly accelerate heavy duty computational tasks in medical physics due to their powerful parallel processing capabilities[25, 26]. We first proposed a GPU-based ultra-fast VMAT optimization method using a column-generation based approach[16]. In this method, the apertures at control points were added sequentially and each newly added aperture was subject to the constraints posted by the already added ones. Dose rate variations were handled in a simple form by a smoothing term in the objective function. Later we further refined this method to include more degrees of freedom and to handle more constraints[18]. Although these methods have been tested in several cases, there are still challenges from the computational point of view. The algorithms have been mainly tested on small, simple prostate VMAT cases with one arc. When it comes to more complicated cases such as head-and-neck (H&N) cancer, the optimization algorithm faces much severer challenges of memory limitation on GPUs, as the existence of multiple targets, much larger target sizes and multiple arcs will substantially increase the size of the DDC matrix. For example, given a CT resolution of 256×256 voxels in a transverse slice, the size of a DDC matrix for a 2-arc H&N VMAT case with three PTVs, stored in a sparse matrix form, is estimated to be ~167GB with a beamlet size $0.5 \times 0.2\ cm^2$, ~67GB with $0.5 \times 0.5\ cm^2$ and ~17GB with $1 \times 1\ cm^2$. However, a GPU card typically has only a few GB memory (up to 12GB for the highest-end GPU card available nowadays), which is much smaller than the size of the DDC matrix obtained even under a coarse resolution. In this situation, the optimization algorithm just fails to launch on a single GPU, as not only the DDC matrix, but also many other temporary variables are needed to be stored. There are some simple ways to overcome the memory limitation on a single GPU. Examples include truncating the DDC matrix to fit the memory of a single GPU, transferring the matrix from CPU to GPU part by part repeatedly for calculation whenever necessary, or moving the calculation part that needs access to the huge DDC matrix from GPU to CPU. Nonetheless, these methods either adversely impact the resulting plan's quality or slow down the computation time, which will be shown in our experimental results.

Another potential way to overcome this memory issue is to utilize a platform with multiple GPUs, as the reduced cost of GPU cards nowadays makes this approach economically affordable. Although the quickly developing GPU technology may increase the memory capacity of a single GPU card and enable us to optimize a large patient case at coarse resolution on a single high-end GPU card, using multi-GPU is better in terms of cost-performance ratio. For example, the cost to build a single-GPU system with one high-end NVIDIA Tesla K40 GPU card (NVIDIA, Santa Clara, CA) is similar to that for a multi-GPU system with two NVIDIA GeForce GTX Titan Z dual GPU cards. But the latter has ~4 times of CUDA cores, 2 times of global memory capacity, ~4.7 times of memory bandwidth and ~3.79 times of peak single/double precision floating point performance. Moreover, the rapid advancement of GPU technology will also evoke our pursuit for solving the VMAT problem at a higher resolution by building a multi-GPU with high-end GPU cards. Hence, it is important to study the multi-GPU implementation of the VMAT optimization problems.

However, since different GPUs only hold their own memory space, parallel processing of this large-scale VMAT optimization problem requires special attention to the associated inter-GPU communications, which should be handled with care to minimize the impact of this data communication overhead on the overall efficiency. It is our main purpose in this paper to present our multi-GPU implementation of the column-generation based VMAT algorithm we have proposed previously. Besides, although we have previously reported our development of the column-generation method for VMAT optimization[16, 18], details about the implementation on the GPU platform have not been discussed. Therefore, another motivation of this paper is to present techniques we utilized to parallelize this algorithm on the GPU platform. In addition, although GPUs have now been employed to solve many medical physics problems, we frequently encounter situations where GPU memory size becomes a limiting factor. To date, studies on multi-GPU solutions are rare. By presenting our multi-GPU implementation in this paper, we also hope to shed some light on this topic using the multi-GPU VMAT optimization as an example problem.

## 2. Methods and Materials

*2.1 Optimization algorithm*

We would like to first briefly describe our column-generation algorithm for VMAT optimization[16, 18]. Let us denote the total number of control points along the treatment arc(s) as $K$. Each control point $k$ is associated with an aperture $A_k$, dose rate $r_k$ (in MU $s^{-1}$), gantry speed $s_k$ (in deg $s^{-1}$), and fluence rate $y_k = r_k/s_k$ (in MU deg$^{-1}$). The beams at all the control points are decomposed into a set of beamlets, denoted by $J$. The patient's CT image is represented by a set of voxels (denoted by $I$). The total dose received by the voxel $i$ ($i \in I$) can be calculated as $z_i = \sum_{k=1}^{K} DA_i(A_k) y_k$, where $DA_i(A_k)$ is the dose contribution from the whole aperture $A_k$ at its unit intensity to the voxel $i$. Our VMAT optimization model employs an objective function with a piecewise quadratic voxel-based penalty:

$$F(z) = \sum_{i \in I} \omega_i^+ \{\max(0,\ z_i - T_i)\}^2 + \omega_i^- \{\max(0,\ T_i - z_i)\}^2, \qquad (1)$$

where $\omega_i^+$ and $\omega_i^-$ are overdose and underdose penalty weighting factors for the voxel $i$, respectively. For PTV voxels, $\omega_i^+$ and $\omega_i^-$ are both positive, whereas for OAR voxels, $\omega_i^+ > 0$ and $\omega_i^- = 0$. $T_i$ is the prescription dose for a PTV voxel and the threshold dose for an OAR voxel $i$.

Our VMAT optimization model can be formulated as

$$\begin{aligned}
\min F(z),\ &\text{subject to} \\
|s_k - s_{k+1}| &\leq \Delta S_k,\ k = 1, \ldots, K-1, \\
s_k &\in [S^L, S^U],\ k = 1, \ldots, K, \\
r_k &\in [R^L, R^U],\ k = 1, \ldots, K, \\
s_k &\leq S^U_{k,k+1}(A_k, A_{k+1}),\ k = 1, \ldots, K, \\
A_k &\in \mathcal{A}_k,\ k = 1, \ldots, K.
\end{aligned} \qquad (2)$$

Here, $S^U$ and $S^L$ are the upper bound and the lower bound on gantry speed, and $R^U$ and $R^L$ denotes the upper bound and the lower bound on dose rate. $\Delta S_k$ denotes the upper bound on the change of gantry speed between control points $k$ and $k+1$. $S^U_{k,k+1}(A_k, A_{k+1})$ denotes the maximum gantry speed that allows MLC to change its shape from $A_k$ to $A_{k+1}$ during the interval between control points $k$ and $k+1$. $\mathcal{A}_k$ is the set of all deliverable apertures for MLC system at control point $k$.

A column-generation approach has been employed here to solve this problem. The algorithm starts with an initial dose distribution $z = 0$ and no apertures. At each iteration step, it attempts to improve the current solution by choosing an aperture among the control points that haven't been occupied yet, and then optimizing the fluence rates of all currently generated apertures. We refer to the problem of selecting a control point and generating a corresponding aperture as a pricing problem (PP), and the problem of optimizing the fluence rates of all the generated apertures as master problem (MP). The algorithm iteratively solves these two problems, until no more apertures at unoccupied beam angles could be found such that the objective function can be further decreased.

In an intermediate iteration step, we denote the control points that have been selected as a set $C \subseteq \{1, \ldots, K\}$ and the corresponding generated apertures as $\{\bar{A}_k, k \in C\}$. The PP can be further divided into two steps. First, we compute the price for each beamlet of all the deliverable apertures $\mathcal{A}_k$ at the unoccupied control points ($k \notin C$). The price for a beamlet $j$ is

$$\pi_j \equiv \sum_{i \in I} g_i(z) D_{ij}, j \in \mathcal{A}_k, k \notin C, \quad (3)$$

$$g_i(z) \equiv -\nabla F(z). \quad (4)$$

$\pi_j$ is essentially the rate of reducing the objective function value in Eq. (1) when the fluence rate of this beamlet is increased by one unit. We would like to point out that if the neighboring control points have been already occupied with chosen apertures, the candidate apertures $\mathcal{A}_k$ need to be compatible with those chosen apertures, which are imposed by the constraints $|s_k - s_{k+1}| \leq \Delta S_k$ and $s_k \leq S^U_{k,k+1}(A_k, A_{k+1})$. Second, based on these calculated beamlet prices, an aperture $\bar{A}_k$ that maximizes the price among all the apertures that satisfy hardware constraints is selected and is added to the aperture set. The DDC vector $DA_i(\bar{A}_k)$ for this chosen aperture is then obtained by summing up the dose deposition vectors for the beamlets within it,

$$DA_i(\bar{A}_k) = \sum_{j \in \bar{A}_k} D_{ij}. \quad (5)$$

With the currently selected apertures and their corresponding $DA(\bar{A}_k), k \in C$, the MP problem is solved to find the optimal fluence rates $y_k$ for them. It can be formulated as

$$\begin{aligned} &\min F(z), \text{ subject to}\\ &z_i = \sum_{k \in C} DA_i(\bar{A}_k) y_k, k \in C,\\ &y_k \in [R^L/S^U_k, R^U/S^L_k]. \end{aligned} \quad (6)$$

Here, $S^U_k$ and $S^L_k$ denote the updated upper bound and the lower bound on gantry speed at the control point $k$, which need to be compatible with the gantry speeds at its neighboring control points.

One modification to the original algorithm developed by Peng *et. al.*[18] is the utilization of Barzilai and Borwein (BB) algorithm[27] with subspace step scheme[28] to solve the MP problem, instead of a gradient descent method combined with a line search for step size. The BB algorithm is able to reach the optimal solution much faster by directly estimating the step size $\delta^{(iter)}$ at each iteration using the following two formulas iteratively,

$$\delta^{(iter)} = \frac{\langle y^{(iter)} - y^{(iter-1)}, y^{(iter)} - y^{(iter-1)} \rangle}{\langle y^{(iter)} - y^{(iter-1)}, g_A^{(iter)} - g_A^{(iter-1)} \rangle}, \quad (7)$$

$$\delta^{(iter)} = \frac{\langle y^{(iter)} - y^{(iter-1)}, g_A^{(iter)} - g_A^{(iter-1)} \rangle}{\langle g_A^{(iter)} - g_A^{(iter-1)}, g_A^{(iter)} - g_A^{(iter-1)} \rangle}. \quad (8)$$

Here, $y^{(iter)}$ denotes the vector of the optimized fluence rates for all the selected apertures at the current iteration. $g_A^{(iter)}$ denotes the gradient of the objective function $F(z)$ in the aperture fluence rate domain. A subspace step scheme is employed to impose the upper and lower bounds onto the fluence rates at each chosen aperture as follows:

If $y_k^{(iter)} \leq R^L/S^U_k$ and $g_{A_k}^{(iter)} < 0$, then $y_k^{(iter)} = R^L/S^U_k$, $g_{A_k}^{(iter)} = 0$, (9)

If $y_k^{(iter)} \geq R^U/S^L_k$ and $g_{A_k}^{(iter)} > 0$, then $y_k^{(iter)} = R^U/S^L_k$, $g_{A_k}^{(iter)} = 0$. (10)

The BB algorithm would be terminated by any one of the three stopping criteria we used, namely, a small enough relative change of the objective function over the last five successive iterations, a small enough step size, or a maximum number of iterations.

Without going into further details, we will move on to present our strategies of implementing this algorithm on a multi-GPU platform. More details about the algorithm can be found in our previous publications[16, 18].

*2.2 multi-GPU implementation*

*2.2.1 multi-GPU system*

Our multi-GPU system is built on a desktop workstation. Two NVIDIA GeForce GTX590 GPU cards are plugged into the motherboard of the workstation. Each of the GTX 590 card contains two identical GPUs, so that there are four GPUs available in our system. These GPUs are labeled as GPU 1 through 4 in the rest of this paper. For each GPU, there are 512 thread processors, each of which has a clock speed of 1.26 GHz. All processors on a GPU share the use of 1.5GB GDRR5 global memory at a 164 GB/sec memory bandwidth. Among GPUs, data transfer is conducted through the computer motherboard via PCIe-16 bus. Our program is written in CUDA 4.0, a C language extension allowing for the programming of each individual GPU, as well as inter-GPU communications.

*2.2.2 multi-GPU data storage*

The largest data set in the problem is the DDC matrix. We will first present the way we store it in the multi-GPU platform, which determines the implementation of our algorithm.

The DDC matrix is a sparse matrix, as each beamlet only has influence on a small fraction of all the voxels. In our system, the DDC matrix is first stored in a coordinate list format (COO) when it is first generated by a finite-size pencil beam dose calculation algorithm[29, 30] via our in-house developed dose engine. This format stores three vectors, namely, a vector consisting of all the non-zero entries of the DDC matrix, a row index vector and a column index vector which denotes the voxel indices and beamlet indices for these non-zero entries, respectively. The lengths of these three vectors are equal to the number of non-zero entries. The DDC matrix is loaded on to CPU and stored in this format, which is to be used later for computing the DDC vector of the chosen aperture, $DA(\bar{A}_k)$, on CPU. The purpose of only storing the COO formatted DDC matrix on CPU is to save the space on GPU for the calculations when solving the optimization problem.

On the GPU side, although Unified Virtual Addressing (UVA) available now could define a universal memory space and thus allow each GPU to access data blindly to the underlying multi-GPU structure, we prefer not to use it in our implementation. This is because under UVA, we do not explicitly control how to split the DDC matrix data among GPUs. During the computations, each GPU thread may access data belonging to other GPUs and the data communication overhead would compromise efficiency. As such, we explicitly split the DDC matrix $D$ into four sub-matrices $D = [D_1, D_2, D_3, D_4]$, where each of the four sub-matrices is a DDC matrix for a group of beam angles. Each GPU then stores one of the four sub-matrices and will be responsible for the calculation of the beamlet price only related to its own DDC sub-matrix (presented later in *2.2.2* session).

Since the DDC matrix is only used on GPU to calculate the beamlet price according to Eq. (3), which can be formally written as $\Pi = D^T g$, only the matrix multiplication of $D_m^T, m = 1,2,...,4$ with a vector will be needed in our algorithm. Hence, we convert $D_m$ into compressed sparse row (CSR) format and store $D_m^T$ on each GPU. This format stores a vector of non-zero elements, a vector of their corresponding column indices, and a vector of pointers for the first elements of each row. We choose this format for two reasons. First, this format is suitable for GPU-based parallel processing of matrix-vector multiplications[31]. Second, because a vector of pointers for the first elements of each row is stored, as opposed to the exact row indices of each matrix element, CSR also requires less memory space. In practice, we found that CSR format uses about only 2/3 memory space compared to the COO format. The conversion of the submatrix $D_m$ into a CSR format is performed at the initialization stage of our algorithm. In addition, there are some other parameters involved in the optimization, such as underdose weighting factors $\omega^+$, overdose weighting factors $\omega^-$, target dose $T$ and structure indices of the voxels, which are all of small data size but high frequently used. Their small data size allows us to store them on all the GPUs in order to avoid some unnecessary data communication between GPUs.

*2.2.2 multi-GPU implementation*

The workflow of our algorithm is shown in Figure 1 (a). In this section, we will present how we implement each of these steps.

Computing beamlet price according to Eq. (3) is the first step, which needs access to the huge DDC matrix and could be parallelized by beamlets. This step is performed on our multi-GPU platform. Since the matrix $D$ has been split into four sub-matrices corresponding to different groups of beam angles, and their transpose $D_m^T$ are stored separately in different GPUs, the computation should be conducted in parallel accordingly. Prior to this step, the current dose distribution $z$ should have been available on GPU 1 from the previous iteration. Therefore, this vector should be sent to all GPUs first. To realize a fast inter-GPU data transfer, peer-to-peer (P2P) memory access and copy is adopted in our implementation, which allows data communication between GPUs directly without going through host memory. To quickly propagate vector $z$ from GPU1 to all other GPUs, we employed a broadcasting scheme shown in Figure 1 (b), where $z$ was first copied from GPU1 to GPU3, and then these two GPUs copied this vector to GPU2 and GPU4, respectively. Note that peer access between participating GPUs (e.g., the access between GPU1 and GPU3, between GPU1 and GPU2 and between GPU3 and GPU4) needs to be enabled at the initialization stage. With this vector $z$ available on all GPUs, we can first compute the gradient vector $g = -\nabla F(z)$ based on Eq. (1). Then the beamlet prices can be computed at each GPU via $\Pi_m = D_m^T g$, which is a simple matrix-vector multiplication. To parallelize this step, a kernel is first launched on all the GPUs to calculate the gradient vector *g* individually on each card, parallelized by voxels with each GPU thread computing one vector component corresponding to the gradient at a given voxel. With the submatrix $D_m$ stored in a CSR format, $\Pi_m = D_m^T g$ can be achieved efficiently on a corresponding GPU using a GPU-based sparse matrix-vector multiplication scheme[31], in which each warp (i.e., a group of 32 threads created, managed, scheduled and executed at a time) is responsible for the price calculation of one beamlet, i.e. the dot product between one row of the sparse matrix $D_m^T$ and the gradient vector *g*. Since warps execute independently, this scheme would greatly alleviate thread divergence, particularly for a matrix with a highly variable number of non-zeros per row. Note that each $\Pi_m$ is a vector of prices calculated for a set of beamlets at beam angles assigned to the $m^{th}$ GPU, therefore we will have to aggregate them to a single GPU for subsequent steps. This is performed in a reduction scheme shown in Figure 1 (b), where the vectors are transferred to GPU1 via a reversed path to what is employed in the previous broadcasting stage.

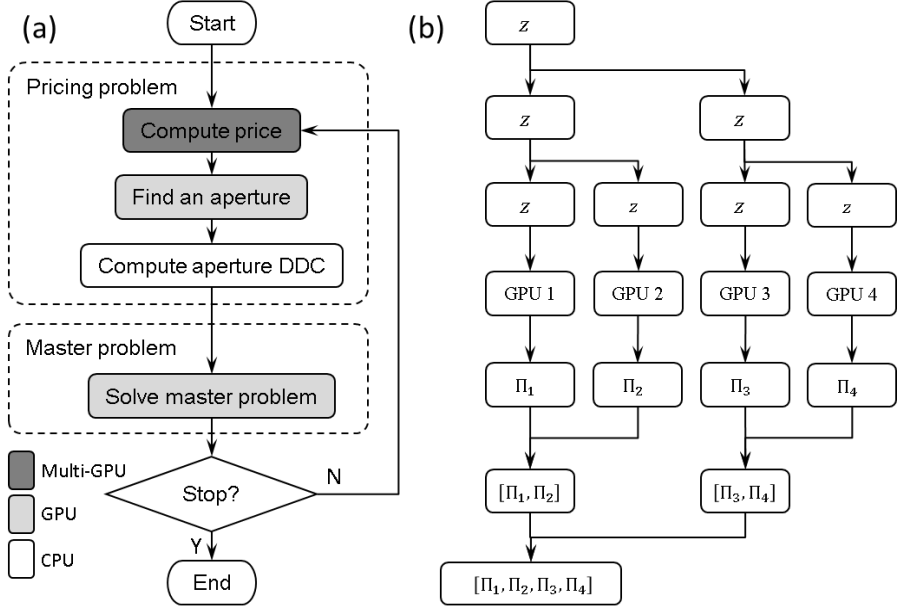

**Figure 1.** (a) Workflow of our algorithm and corresponding platform for each key step. (b) Illustration of broadcast and reduction scheme on multi-GPUs when computing the beamlet price.

The next step in the PP problem is to find a feasible aperture at an unoccupied control point that encompasses an area with the best price. Since this step doesn't need an access to the huge DDC matrix and has modest amount of computation load, parallelized by MLC rows/apertures, we choose to implement it on GPU1. Two kernels have been developed to achieve this. In the case of not using interdigitation constraint, MLC rows are independent of each other and hence this step of the PP problem can be solved row by row. Therefore, the first kernel is to find the best opening $(L_{k,irow}, R_{k,irow})$ among all candidate openings for each MLC row at all the unoccupied control points, with each thread responsible for one row. Here, $L_{k,irow}$ denotes the position of the MLC left leaf on the $irow_{th}$ row for the $k_{th}$ control point, while $R_{k,irow}$ denotes the position of the corresponding MLC right leaf. After that, another kernel is launched to let each thread calculate the best price for the aperture at each unoccupied control point. The aperture with the best price is chosen as $\bar{A}_k$. When there is an interdigitation constraint of MLC leaves in adjacent rows, that is, $L_{k,irow} \leq \min\{R_{k,irow-1}, R_{k,irow+1}\}$, the MLC rows are not independent any longer and we cannot decompose the problem of finding the best aperture by row. In this case, each thread in the first kernel is only responsible for calculating the prices of all candidate openings for one MLC row. Then in the following kernel, each thread would be responsible for one control point and for calculating the prices of all the combinations of openings at different MLC rows that can satisfy the interdigitation constraint. It would then choose the best aperture for that control point.

Once a new aperture $\bar{A}_k$ is selected, its corresponding DDC vector $DA(\bar{A}_k)$ is calculated, which is necessary for the subsequent MP problem. $DA(\bar{A}_k)$ is a vector that specifies the amount of dose exposed to each voxel by this aperture per unit intensity. It can be calculated by summing up all the columns in the full DDC matrix $D$ that correspond to the beamlets within this aperture. Since the full matrix is stored only on CPU in a COO format to save memory on GPU, we perform this computation on the CPU platform. Efficiency is not a concern at this step, as its computation load is relatively light. In addition, due to the CSR storage format of $D$ among GPUs, this step couldn't be computed on GPU, unless an additional vector of the row indices (the voxel indices) is also transferred onto GPU, which would inevitably increase the required memory space on GPUs.

Finally, once the DDC vectors for all the chosen apertures $DA(\bar{A}_k), k \in C$ are available, the next step solves the MP problem specified in Eq. (6), which is essentially a least-square minimization problem. The BB algorithm[27,28] employed here to solve the MP problem involves four main steps, namely calculating the gradient of the objective function in the aperture fluence rate domain, estimating the step size using Eq.(7) and (8) alternately, updating the fluence rate of apertures within the subspace according to Eq.(9) and (10), and then updating the dose distribution accordingly. These four steps are iterated to find the optimal fluence rate for all the chosen apertures. Although the scale of the MP problem increases during the iterative process, as more and more apertures are generated, the overall problem scale of MP is still quite modest. This problem is hence determined to be implemented on GPU1 only to take advantage of single GPU parallelization. The parallelization of the step 1 and 3 in the BB algorithm is very straightforward, as each thread could be responsible for one chosen aperture. The step 4 is parallelized with each thread updating one voxel. The NVIDIA CUDA basic linear algebra subroutines (cuBLAS) library[32] is adopted in step 2 to calculate the step size at each iteration.

## 2.4 Materials

To demonstrate the feasibility of this multi-GPU implementation and its advantages over other possible solutions on a single GPU for speed and memory limitations, we first study a H&N patient case, which has three PTVs and their prescription dose are 70 Gy, 59.4 Gy and 54 Gy, respectively. Two coplanar arcs with 354 equal-spatially distributed control points are used for optimization. Beamlet size is set to be $1 \times 1$ cm$^2$ and voxel size is $0.195 \times 0.195 \times 0.25$ cm$^3$, namely, 256×256 resolution on axial slices. DDC matrix for this case is computed using our pencil-beam dose engine[30]. The DDC matrix for this problem occupies 6.80 GB memory space in COO format and 4.53 GB in CSR format. We would like to point out that this matrix size is already after downsampling the patient CT image in order to fit the matrix in our multi-GPU system for demonstration purpose.

Despite this huge DDC matrix, there are still some possible solutions to realize VMAT optimization on a single GPU. We have also implemented three of them for the purpose of demonstrating advantages of our multi-GPU method. In the first method (S1), the huge DDC matrix was truncated to make it possible to perform the VMAT optimization on a single GPU. Specifically, we used a truncation threshold γ, the radius at isocenter level of a circular cone centered at each beamlet's central axis. In the DDC matrix, those elements corresponding to the voxels outside this cone were truncated. The smaller γ was, the more aggressive the truncation was. We used a large enough γ=6 cm to obtain a complete DDC matrix, and γ was set to 2.5 cm to truncate the DDC matrix for S1. In the second method (S2), we divided the DDC matrix into several small parts. During the optimization, the matrix was transferred from CPU to GPU part by part whenever the DDC matrix was needed to calculate the prices of the corresponding beamlets. In the third one (S3), we moved the calculations of beamlet price, where the access to DDC matrix was needed, onto CPU. The plan quality and computation time for this H&N patient case were compared among our multi-GPU implementation and these three single-GPU implementations. For clarity, we refer the plan obtained on our multi-GPU system to Plan-M, and denote the plans obtained on a single GPU with these three strategies as Plan-S1, Plan-S2, Plan-S3, respectively.

In addition, we have studied five more realistic clinical cases, including two H&N patient cases and three prostate cases with the size of DDC matrix (in COO format) ranging from 2.53 GB to 4.77 GB to demonstrate that we can consistently achieve a high computational efficiency for the VMAT optimization problem. The relatively small target size of prostate cases allows us to use a $0.5 \times 0.5$ cm$^2$ beamlet size. We also optimized these VMAT cases using Eclipse treatment planning system (Varian, Palo Alto, CA) installed on an Intel Xeon E5620 CPU card@2.4GHz to achieve clinically acceptable and comparable treatment plans and recorded the optimization time for comparison.

## 3. Experimental Results

*3.1 plan quality*

The Dose-volume histograms (DVH) of the resulting plans are shown in Figure 2. A set of reference dose/volume limits is also used to quantitatively evaluate the plan quality, and the results are listed in Table 1. Note that our multi-GPU implementation and the single GPU implementation S2 and S3 should give the same plan, whereas S1 should give an inferior plan due to the truncation in the DDC matrix. The solid lines shown in Figure 2(a) and (b) represent the dose distribution of Plan-M (and hence Plan-S2 and Plan-S3). Combined with the results in Table 1, we can see that our multi-GPU-based VMAT optimization algorithm achieves a good plan for this H&N patient case. The dashed lines in Figure 1(a) depict DVHs of Plan-S1's optimized dose. These DVHs demonstrate that the optimization performs well on the single GPU using this truncated DDC matrix, and the DVH curves of three PTVs match those of Plan-M. It seems that OARs are better spared in Plan-S1 than in Plan-M, since the truncation of the DDC matrix makes the optimization problem easier. However, by adding the dose distribution corresponding to the deleted small elements of the DDC matrix, which are ignored in optimization, the actual dose of Plan-S1 represented by the dashed lines in Figure 2(b) becomes much higher than the dose of Plan-M. It was observed that this matrix truncation resulted in ~8% dose difference of PTV dose. According to Table 1, more than 10% of the volume receives a dose higher than 110% of the prescription dose for each PTV. The reason of this big dose discrepancy in PTVs is because each beamlet involved in the optimization goes through or is laterally very close to the PTVs. Although only penumbra elements for each beamlet dose were deleted, the accumulated dose contributed from all the deleted elements to a voxel inside PTV could still be relatively large. We would like to point out that there are algorithms developed to reduce the DDC matrix size, while better preserving the plan quality[33]. Here, we utilized a simple method for the purpose of focusing on the study about multi-GPU implementation. Moreover, the maximal doses received by spinal cord and both parotids are also found to be above their thresholds.

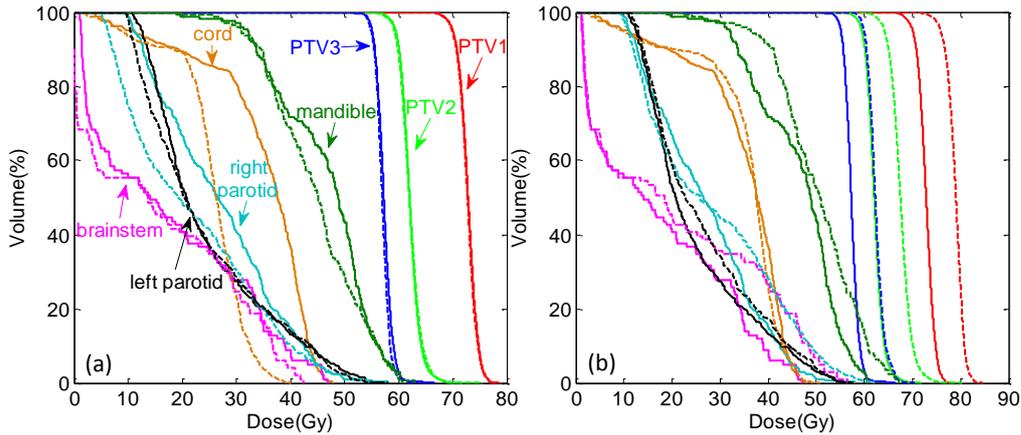

**Figure 2.** Dose-volume histograms of the resulting plans. (a) Solid lines depict Plan-M (Plan-S2 and Plan-S3); dashed lines illustrate the optimized major dose of Plan-S1; (b) Solid lines depict Plan-M as well for comparison purpose; dashed lines illustrate the actual dose of Plan-S1. Same colors as in (a) are used in (b) for different PTVs and OARs.

**Table 1.** Plan quality evaluation.

| Structure | Dose (Gy) | Criterion | Plan-M, S2, S3 | Plan-S1 |
|---|---|---|---|---|
| | 65.1 | ≥99% | 99.96% | 100% |
| PTV1 | 70.0 | ≥95% | 95.24% | 100% |
| | 77.0 | ≤10% | 0.40% | 87.94% |
| | 55.2 | ≥99% | 99.90% | 100% |
| PTV2 | 59.4 | ≥95% | 95.30% | 99.95% |
| | 65.4 | ≤10% | 3.65% | 86.51% |
| | 50.2 | ≥99% | 99.88% | 100% |
| PTV3 | 54.0 | ≥95% | 95.88% | 100% |
| | 59.4 | ≤10% | 1.34% | 89.02% |
| brainstem | 60 | ≤1% | 0% | 0% |
| Mandible | 75 | ≤1% | 0% | 0% |
| Spinal cord | 48 | Max | 47.92Gy | 50.79 Gy |
| Parotid-both | 26 | Mean | 25.79Gy | 27.50 Gy |

*3.2 Computation time*

The size of the complete DDC matrix is about 6.80 GB in COO format and 4.53 GB in CSR format. In contrast, the size of the truncated DDC matrix for S1 on single GPU is about 1.79 GB in COO format and 1.19 GB in CSR format. The number of beamlets and voxels involved in the DDC matrix and the number of non-zero elements in the DDC matrix are listed in Table 2. The computation times for different implementation strategies are shown in Table 2. Here, $T_{total}$ denotes the total computation time needed to get a plan. The times of some steps are also listed, including 1) $T_{data-read}$ that represents the time spent on loading the input data from disk onto CPU memory; 2) $T_{data-transfer}$ that denotes the time spent on transferring the data from CPU to GPU. Note that for multi-GPU implementation, $T_{data-transfer}$ also includes the time for inter-GPU data transfer; 3) $T_{PP1}$ that denotes the time for a substep of the PP problem, namely, computing the prices of all beamlets (referred as PP1), where the DDC matrix is involved; 4) $T_{PP2}$, the time for the rest of the PP problem, namely, finding an good aperture based on current solution and calculating the DDC vector for the chosen aperture (referred as PP2); 5) $T_{MP}$, the time for the MP problem; 6) $T_{others}$, the time for all other steps, e.g. memory allocation, memory free, outputting results onto files, etc. The number of iterations that is needed in our VMAT optimization algorithm to reach a good plan for this H&N patient case and the number of inner iterations of the BB algorithm used to solve the MP problem are shown in Table 2 as well.

From Table 2, we can see that our multi-GPU-based VMAT optimization takes about 1 minute to reach a solution. Specifically, over 16 seconds are spent for loading such a huge DDC matrix and other inputs from disk onto CPU memory. The time is the same for S2 and S3, due to the same DDC matrix size, and is shorter for S1 where the DDC matrix was truncated. For our multi-GPU implementation, about 9 seconds are used to transfer these data from CPU to the four GPUs, with each GPU storing a DDC submatrix for a quarter of the control points. Note that the inter-GPU communications during the optimization, illustrated in Figure 1(b), is not a concern for our application, taking only 40 milliseconds in total. This might be explained by two reasons. First, the inter-GPU data communication occurs only when our VMAT optimization algorithm iterates between MP and PP, and the number of the iterations is relatively small, e.g. 199 iterations for this H&N case. Second, direct P2P data communication without going through host memory is expected to have high bandwidth and low latency.

For Plan-S1, obtained on a single GPU by strategy 1, the total time was about 14 seconds shorter than that for the Plan-M, as the time spent on data reading and transferring became shorter and the computation load was lighter, both due to the truncation of the DDC matrix. However, the plan quality is inferior and not acceptable as illustrated before in Figure 2 and Table 1. In addition, since the calculation of beamlet price is implemented on single GPU, even with the relatively small truncated DDC matrix, the time for this step is still about 1.4 times as long as the corresponding time for multi-GPU implementation. For Plan-S2, obtained on a single GPU by strategy 2 with the complete DDC matrix, the total computation time is 255 seconds, about four times as long as the time of our multi-GPU implementation. Specifically, it takes about three minutes in total to transfer the DDC matrix from CPU to GPU part by part to calculate the beamlet price whenever needed, which accounts for about 2/3 of the total time. The time for the calculation of the beamlet price on a single GPU is about four times of that in the multi-GPU implementation. This reflects the net gain of utilizing a multi-GPU platform to speed up computations. In strategy 3, we try to avoid the memory issue of a single GPU as well as long data transfer time from CPU to GPU by directly calculating the beamlet price on CPU. The pthread library was used to parallel this calculation part on CPU side using all the four cores and eight threads for fair comparison. This strategy turns out to be the slowest scenario with a total computation time of 6 minutes, which is as ~5.6 times long as our multi-GPU implementation. In addition, the computation time for the substep PP1 on CPU was ~8.4 times longer, compared to the calculation implemented on a single GPU in S2. Regarding the time for other steps, such as $T_{PP2}$, $T_{MP}$ and $T_{others}$, the computation time is more or less the same among different implementation schemes since these steps are implemented in the same fashion, except for S1. S1 has longer $T_{PP2}$ due to the more iterations taken in optimization and shorter $T_{MP}$ mainly due to its smaller DDC matrix. These results have shown that our multi-GPU strategy has shortest computation time among different strategies without compromising plan quality.

**Table 2.** Computation times of different implementation strategies.

|  |  | Plan-M | Plan-S1 | Plan-S2 | Plan-S3 |
|---|---|---|---|---|---|
| DDC matrix size (GB) | COO | 6.80 | 1.79 | 6.80 | 6.80 |
|  | CSR | 4.53 | 1.19 | 4.53 | 4.53 |
| Number of beamlets |  | 128516 | 128516 | 128516 | 128516 |
| Number of voxels |  | 26828 | 26828 | 26828 | 26828 |
| Number of non-zero dij elements |  | 608447419 | 159839137 | 608447419 | 608447419 |
| Number of iterations (apertures) |  | 199 | 245 | 199 | 199 |
| Number of BB inner iterations |  | 9364 | 11529 | 9364 | 9364 |
| Number of GPU cards |  | 4 | 1 | 1 | 1 |
| Time (s) | $T_{total}$ | 63.96 | 50.46 | 255.32 | 360.85 |
|  | $T_{data-read}$ | 16.51 | 9.10 | 16.70 | 16.65 |
|  | $T_{data-transfer}$ | 9.30 (9.26 + 0.04) | 7.11 | 175.21 | 0.01 |
|  | $T_{PP1}$ | 10.77 | 14.92 | 38.33 | 320.37 |
|  | $T_{PP2}$ | 5.05 | 6.31 | 5.16 | 4.99 |
|  | $T_{MP}$ | 21.24 | 12.47 | 19.33 | 18.30 |
|  | $T_{others}$ | 0.64 | 0.55 | 0.59 | 0.53 |

In addition, it was observed that the PP2 substep, which was partially implemented on a single GPU to find the best aperture and partially on CPU to calculate the DDC vector of the chosen aperture in all the four implementation strategies, only takes about 5~6 seconds. Even in an ideal situation with a speed-up factor of four under the multi-GPU system, we could only save 3~4 seconds. This gain on computation time would be even less considering communication overhead. This is the reason why we didn't spend a lot of effort to move this part from primary GPU onto multi-GPU.

Conversely, the time on MP problem was about 21 seconds, accounting for ~33% of the whole optimization process in the multi-GPU version. Due to this relative large time portion, we did some experiments to implement the MP problem on multiple GPUs in order to see whether we could achieve further efficiency gain. The computation time of MP on multiple GPUs turned out to be 22.98 s, compared to 21.24 s when only using the primary GPU. This implies that the inter-GPU and GPU-CPU data communication overhead counteracted the small computational efficiency gain obtained by utilizing multiple GPUs, which justifies our strategy to implement the MP problem only on the primary GPU.

*3.3 GPU performance analysis*

For better comparison between our multi-GPU implementation and the three single GPU implementation strategies, the GPU performance in terms of GPU occupancy and device memory usage was analyzed. The major difference between these different implementations was where to calculate the beamlet price, i.e. on multiple GPUs, on a single GPU or on CPU and whether to truncate the DDC matrix for calculation or not. Hence, we first analyzed the GPU occupancy of the beamlet price calculation kernel implemented either on multiple GPUs in our multi-GPU implementation or on a single GPU in our single-GPU implementation strategies S1 and S2. It was found that these three different implementations turned out to have same GPU occupancy. Specifically, there were 1536 active threads, 48 active warps and 3 active thread blocks per multiprocessor. The occupancy of each multiprocessor was 100% no matter whether the kernel was implemented utilizing all the four GPUs or only a single GPU. The reason was that the multi-GPU version of this kernel and the single GPU version for S2 were actually the same, which only required two more registers per thread to indicate the data arrangement of the DDC submatrices for calculation, compared to the single GPU version used for S1. Given 18 registers per thread used in the kernel for S1, the usage of two more registers per thread didn't impair the GPU occupancy.

Second, the information on device memory usage was acquired for the whole VMAT optimization process for our multi-GPU implementation and all the three single GPU implementation strategies. For our multi-GPU implementation, the percentage of the used device global memory relative to the total device memory of one GPU (i.e., 1.5GB) was about 83.48%, 81.87%, 81.72% and 80.50% for each GPU, respectively. It was noticed that the mount of memory usage on the primary GPU, namely GPU1, was the largest, since PP2 and MP were implemented only on this GPU and hence more storage space was required. For the single-GPU implementation strategy S1, which truncated the DDC matrix to fit into a single GPU's memory, ~87.43% of the device global memory was used. For the single GPU implementation strategy S2 that transferred the complete DDC matrix from CPU to GPU part by part for calculation whenever needed, ~83.48% of the device global memory was used, which was the same to the memory usage on the primary GPU in multi-GPU implementation. For strategy S3, which moved the calculation of beamlet price onto CPU, only ~8.02% of the device memory was used. The analysis results of GPU performance illustrated that multi-GPU system could help us to overcome the GPU memory issue without impairing the GPU occupancy.

*3.4 Validation on other cases*

The total optimization time of our multi-GPU VMAT optimization for the H&N patient case tested before (denoted with *) and 5 other patient cases (two H&N and three prostate cases) are listed in Table 3. Some plan specification, such as the number of arcs $N_{arc}$, the number of control points $N_{cp}$, the number of PTVs $N_{PTV}$, the number of involved OARs $N_{OAR}$ and beamlet size are also listed in the table, as well as the sizes of the DDC matrices in both COO and CSR format. The DVH curves of H&N patient 2 and prostate patient 1 are shown in Figure 3. These VMAT cases were also optimized using Eclipse TPS to get clinically acceptable and comparable treatment plans, and the corresponding optimization time and DVH curves are also presented in Table 3 and Figure 3 for comparison.

Again, it can be seen from the Table 3 that the CSR format uses only 2/3 memory space compared to the COO format. We would like to point out that since all three H&N cases in this study have three PTVs and are much larger than the target size of the prostate cases, we have to use 1×1 cm² beamlet size for the H&N cases in order to fit into the memory of our multi-GPU system. Conversely, having only one PTV of relatively small size, those prostate cases allow us to use a higher resolution of beamlets, i.e. $0.5 \times 0.5$ cm² beamlet size, for VMAT optimization. Although the complete DDC matrices of these cases are all too large to be stored in a single GPU on our multi-GPU system to launch the optimization, we have consistently achieved high computational efficiency (24~64 s) in solving these large-scale VMAT optimization problems on our multi-GPU platform. The qualities of the obtained VMAT plans for these cases are clinically acceptable, and the DVH curves of H&N patient 2 and prostate patient 1, as shown in Figure 3, demonstrate again the efficacy of our multi-GPU based VMAT optimization. Note that the plan quality is related to the underdose and overdose weighting factors we set for PTVs and OARs and might be improved further by fine-tuning these parameters. However, this is not a focus of our paper.

In the Eclipse TPS system installed on an Intel Xeon CPU, it took about 233~710 seconds to achieve clinically acceptable and comparable treatment plans for these VMAT cases, which is 5.8~14.0 times slower than our multi-GPU based optimization approach. Here, we would like to point out that this comparison might not be completely fair due to the different optimization models. Besides, since we have to manually adjust both the weighting factors in our algorithm and the dose constraints and priorities in the commercial TPS system to tune the resulting plans' quality, it is probably not possible to get two sets of identical plans. Hence, we tried to make both of the two sets of plans clinically acceptable and comparable in quality. Although the specific optimization time was impacted by the different trade-off among PTV and OARs, imposed by the dose constraints in the optimization algorithms, the order of magnitude of the optimization time remained unchanged. Specifically, the optimization time needed in Eclipse TPS system was found to be several minutes and the optimization time on our multi-GPU system was kept within about 1 minute. Therefore, the comparison between these two algorithms on different platforms illustrates the efficiency advantage of our algorithm on the multi-GPU system to a certain extent.

**Table 3.** Plan Info and Optimization time of our algorithm on Multi-GPU system and the Optimization time of Eclipse TPS system.

|  | $N_{arc}$ | $N_{cp}$ | $N_{PTV}$ | $N_{OAR}$ | Beamlet size (cm$^2$) | DDC size (GB) COO | DDC size (GB) CSR | Multi-GPU Opt time (s) | Eclipse Opt time (s) |
|---|---|---|---|---|---|---|---|---|---|
| H&N P1* | 2 | 356 | 3 | 5 | 1×1 | 6.80 | 4.53 | 63.96 | 710.00 |
| H&N P2 | 2 | 356 | 3 | 5 | 1×1 | 4.23 | 2.82 | 31.50 | 440.00 |
| H&N P3 | 2 | 356 | 3 | 4 | 1×1 | 4.62 | 3.08 | 37.36 | 295.00 |
| Prostate P1 | 2 | 356 | 1 | 4 | 0.5×0.5 | 4.77 | 3.18 | 39.77 | 312.00 |
| Prostate P2 | 2 | 356 | 1 | 2 | 0.5×0.5 | 4.59 | 3.06 | 45.46 | 261.00 |
| Prostate P3 | 1 | 178 | 1 | 4 | 0.5×0.5 | 2.53 | 1.69 | 23.28 | 233.00 |

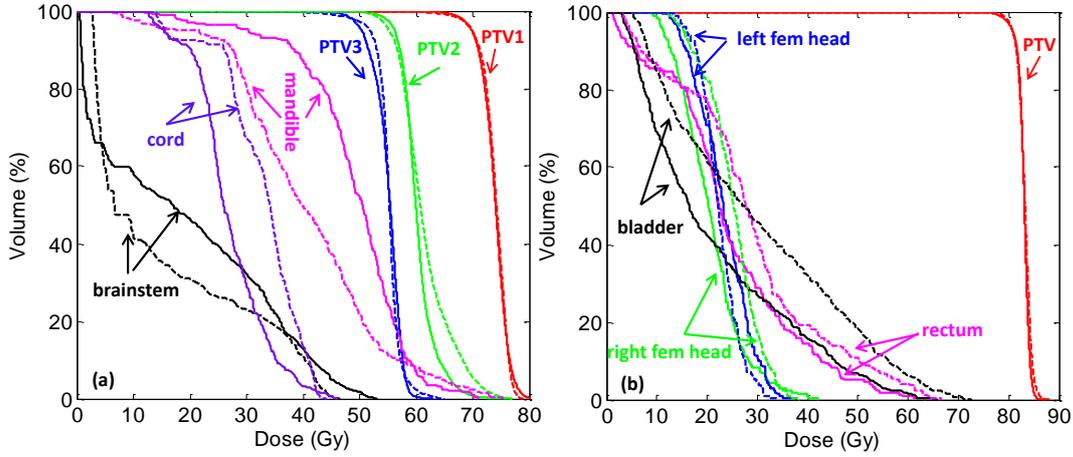

**Figure 3.** DVH curves of the VMAT plan obtained for H&N patient 2 (left) and prostate patient 1 (right). Solid lines denote the plan obtained from our multi-GPU VMAT optimization algorithm; Dashed lines denote the plan obtained from Eclipse TPS system. Note that for the H&N case, the DVH curves of left and right parotids involved in optimization are not shown here for clarity.

## 4. Discussion and conclusions

VMAT treatment planning is usually a large-scale optimization problem and subjects to many coupling hardware constraints, which makes it computationally challenging. GPU's powerful parallel computation ability can help to speed up the optimization. However, its very limited memory makes it difficult to handle a huge DDC matrix. In light of the affordable cost of GPU cards nowadays, we have developed a multi-GPU implementation for VMAT optimization. Our VMAT optimization algorithm is based on a column-generation approach, which generates apertures one by one by solving two sub-problems, PP and MP, iteratively. In our implementation, the full DDC matrix in COO format is only stored on CPU for aperture DDC vector calculation on the CPU side in order to save memory on GPU. The huge DDC matrix is handled on multi-GPUs by dividing it into four sub-matrices and each GPU stores one sub-matrix in CSR format, which saves about 1/3 memory compared with the COO format. The first step of PP, namely to calculate the beamlet price, needs an access to the huge DDC matrix, and is thus implemented on multi-GPUs. Broadcast and parallel reduction schemes are adopted for data transfer and communication between GPUs. The remaining step of PP and the MP are implemented on CPU or single GPU due to their modest amount of computational load. We compared the results of our multi-GPU scheme with three different strategies implemented on a single GPU by truncating the DDC matrix, transferring the DDC matrix from CPU to GPU part by part, or moving this calculation onto CPU, respectively. Although strategy 1 on single GPU with the truncated DDC matrix saves us about 10 seconds compared with the multi-GPU scheme, the actual plan-S1 turns out to be an inferior one. Due to the complete DDC matrix used for optimization, the strategies 2 and 3 implemented on single GPU generate the same plan as the multi-GPU scheme did. However, the computation time for strategy 2 and strategy 3 are much longer than our multi-GPU implementation. We have also tested our method in 2 more H&N patient cases and 3 prostate patient cases and compared with the optimization method in the commercial Eclipse TPS system. It was found that we can consistently achieve a high computational efficiency (24~64s) for the VMAT optimization problem on the multi-GPU platform. The corresponding optimization time for these VMAT cases in the commercial TPS on CPU platform was found to have an order of magnitude of several minutes, which also demonstrated the efficiency gain of multi-GPU solutions of this large-scale VMAT optimization problem.

It is possible that all the cases shown in the paper could be fit into a single high-end GPU card now or in near future. However, we would like to point out that we have tailored these cases using a coarse resolution to make the DDC matrices already small. A multi-GPU system with relative low-end GPU cards usually has much better cost-performance tradeoff than a single-GPU system with a high-end GPU card, making the multi-GPU approach attractive. Moreover, given high-end GPU cards with large memory sizes, the multi-GPU solution will enable us to solve the VMAT optimization problem at a higher resolution.

There are some other methods that can be used to alleviate the GPU memory issue. For example, the importance sampling of the DDC matrix was reported to be able to reduce the matrix size to be ~1/3 of the complete matrix without sacrificing plan quality[33]. These methods could be combined with our multi-GPU implementation to allow us to keep DDC elements as many as possible or use even higher resolution in VMAT optimization on GPUs.

Regarding the suitability of the VMAT problem for multi-GPU parallelization, we have analyzed the algorithm structure. First, for PP1, the substep of PP to calculate the beamlet price, the problem scale is proportional to the data size of the DDC matrix. This part is parallelizable by distributing the computations for different sub-DDC matrices to different GPUs and hence potentially benefits the most from multi-GPU implementation. Meanwhile, the percentage of computation load for this step over the entire algorithm depends on the problem size, e.g. voxel resolution. For the large cases, this part dominates and is hence the most needed for multi-GPU parallelization. Second, the computation time for PP2 is usually a very small portion of the whole process, so the gain in computation time from multi-GPU implementation of this step would be negligible. This is the reason why we didn't spend a lot of effort to move this part from primary GPU onto multi-GPU. Third, for the MP problem that only handles the DDC for currently chosen apertures at each iteration, the data involved in this problem is much smaller than that in PP1. For example, as the number of apertures increased from zero to 199 during the optimization for the H&N patient case1, the largest data size in MP occurred at the last iteration is about 0.15% of the DDC matrix size used in PP1. Due to its modest scale, multi-GPU parallelization didn't speed up MP further, as the inter-GPU and CPU-GPU data communication overhead counteracted the small efficiency gain of utilizing multiple GPUs. Single GPU parallelization turns out to be a better option for MP in terms of efficiency, memory, and the code's simplicity as well. Based on these observations, the overall performance of our VMAT optimization algorithm may not scale well on multi-GPUs in the cases tested. Since it is not possible to run the HN case on a single GPU due to the small memory size of our current card, we tried to predict the computation time as following. Based on the column Plan-M in Table 2, if we were to run the problem on a single GPU, $T_{PP1}$ would become 3 times longer, namely ~43 sec. At the same time, inter-GPU communication can be avoided and hence we gain 0.04 sec. Note the CPU-GPU communication time of 9.26 sec still exists. This leads to an estimated total computation time of ~96 sec on a single GPU. So the gain using multi-GPU in this case is not quite significant in terms of efficiency (~64 sec versus ~96 sec). This is in fact due to two factors: 1) the nature of the problem, where PP2 and MP is not well suited for multi-GPU parallelization and 2) the data loading overhead (16.5 sec from hard drive to CPU, and 9.3 sec from CPU to GPU) that cannot be avoided. However, our multi-GPU strategy does have its own value. 1) The large problem size of VMAT for large cases prevents it from running in a single GPU. Using multi-GPU makes this computation possible. 2) The computation time achieved using our multi-GPU strategy is by far the shortest among the four different strategies without compromising plan quality. 3) It is expected that for a large case, e.g. with a large tumor size and/or under a high resolution, the portion of PP1 will be more and more significant, which makes the entire VMAT algorithm more suitable for multi-GPU parallelization.

The multi-GPU system probably does not utilize GPU 2~4 optimally for our VMAT optimization application, as they only work in the PP problem at each iteration step, when access to the huge DDC matrix is needed. Moreover, in our multi-GPU-based VMAT optimization, the time spent for MP problem implemented on a single GPU is ~1.3 times as long as the time for PP problem. A scheme to better use the multi-GPU system might be to flexibly determine the number of needed GPUs based on the size of the DDC matrix for different cases, and make the unused GPUs available to some other applications at the same time, such as another VMAT planning for small cases or IMRT optimization. These will be our future studies.

We also hope this study would shed some light on other problems that require multiple GPUs. One of the drawbacks of GPU is its limited memory space. This is, and will probably continue to be, a big concern when utilizing this novel platform, in light of the increasing size of a variety of problems in medical physics. Multi-GPU is a practical approach to overcome this problem given the low cost of GPU cards nowadays. However, parallelization scheme should be carefully designed to ensure performance. In addition to VMAT optimization problems, other inverse planning problems involving huge DDC matrices may benefit from the implementation here. Examples include beam orientation optimization[34-36] and 4 PI treatment planning[37-39]. Another group of problems that benefits from GPU accelerations but is limited by its memory size is iterative cone beam CT (CBCT) reconstruction[40, 41], especially 4D CBCT problems[42-44] where the additional temporal dimension substantially increases the problem size. The fundamental operations for CBCT reconstructions are solving a linear equation corresponding to the x-ray projection. Matrix-vector operations are the most frequent operations in the linear equation. It is expected that the parallelization scheme in this paper could be of help for developing multi-GPU-based iterative CBCT reconstruction algorithms.

**Acknowledgement**

The authors would like to thank Ms. Sarah Sandle for proof-reading this manuscript.